\documentclass[a4paper,12pt]{article}
\pdfoutput=1
\usepackage{cite}
\usepackage{amsmath}
\usepackage{amsfonts}
\usepackage{amssymb}
\usepackage{graphicx, rotating}
\usepackage{epsfig}
\usepackage{latexsym}
\usepackage{graphicx}
\usepackage{color}
\usepackage{amsmath,bm,amssymb}
\usepackage{booktabs}
\usepackage{geometry}

\usepackage{color}
\usepackage[english]{babel}
\usepackage{hyperref}

\newcommand{\Slash}[1]{{\ooalign{\hfil#1\hfil\crcr\raise.167ex\hbox{/}}}}

\newcommand{\beq}{\begin{equation}}  \newcommand{\eeq}{\end{equation}}
\newcommand{\bef}{\begin{figure}}  \newcommand{\eef}{\end{figure}}
\newcommand{\bec}{\begin{center}}  \newcommand{\eec}{\end{center}}
  
\newcommand{\laq}[1]{\label{eq:#1}}

\newcommand{\eq}[1]{(\ref{eq:#1})}
\newcommand{\Sec}[1]{Sec.\ref{chap:#1}}

\newcommand{\lac}[1]{\label{chap:#1}}

\def\({\left(}
\def\){\right)}

\def\O{\mathcal{O}}

\newcommand{\KEV}{ {\rm \, keV} }

\newcommand{\GEV}{ {\rm \, GeV} }

\def\f{\phi}
\def\g{\gamma}

\def\l{\lambda}

\def\s{\sigma}

\def\D{\Delta}
\def\G{\Gamma}
\def\F{\Phi}

\def\*{\dagger}

\begin{document}
\renewcommand\bibname{\Large References}

\begin{center}

\vspace{1.0cm}

{\Large\bf Double Narrow-Line Signatures of Dark Matter Decay and New Constraints from XRISM Observations }

\vspace{1.0cm}

{\bf  Wen Yin, Yutaka Fujita, Yuichiro Ezoe, and Yoshitaka Ishisaki}

\vspace{0.5cm}
{\em 
{Department of Physics, Tokyo Metropolitan University, Tokyo 192-0397, Japan} 
}

\vspace{1cm}

\begin{abstract}
We investigate the indirect detection search of the two-body decay of dark matter particles into final states containing a  photon, a process predicted in various promising dark matter models such as axion-like particles and sterile neutrinos. Recent and near-future photon detectors with a resolution 
$
R \equiv \lambda/\Delta\lambda =  \O(1000)
$
 are primarily optimized for the velocity dispersion of dark matter in the Milky Way. When performing indirect detection of dark matter in objects other than the Milky Way, one should take into account the contribution from Milky Way dark matter. As a result, the dark matter signal observed by a detector is predicted to exhibit a two-peak structure in many targets, owing to the Doppler shift, differences in radial velocities and the good energy resolution. An analysis incorporating this two-peak effect was performed using the latest XRISM observation data of the Centaurus galaxy cluster~\cite{XRISM:2025axf}.
Although, due to the relatively short observation time, our derived limit is weaker than some existing limits, among dark matter searches in galaxy clusters our limit is one of the most stringent (at least in certain mass ranges). 
We also perform the usual single-peak analysis, for considering the various scenarios, that prefer narrow-line photon from the faraway galaxy cluster. Future data releases from XRISM as well as other observatories will further strengthen our conclusions.
\end{abstract}

\end{center}
\clearpage

\setcounter{page}{1}
\setcounter{footnote}{0}


\section{Introduction}
The fundamental nature of dark matter remains one of the most pressing open questions in particle theory, astrophysics and cosmology. Among the plethora of candidate models, those in which dark matter particles decay or annihilate into Standard Model particles have attracted considerable attention. In particular, the two-body decay of dark matter particles into photon-emitting final states is a promising avenue for indirect detection (e.g.,~\cite{Gruber:1999yr,Bouchet:2008rp,kappadath1998measurement,Strong:2004de,Fermi-LAT:2012edv,Horiuchi:2013noa,Tamura:2014mta,Riemer-Sorensen:2015kqa,Cirelli:2020bpc,Foster:2021ngm,2024PASJ76512F}), as it yields a spectral line whose characteristics can be directly linked to the underlying particle properties. Indeed, indirect detection is one of the most powerful tools for probing particle models such as sterile neutrinos, axion-like particles (ALPs), or light moduli (see reviews, e.g.,~\cite{Abazajian:2012ys,Boyarsky:2009ix,Drewes:2016upu,Abazajian:2017tcc,Acero:2022wqg} for sterile neutrinos, and ALPs~\cite{Jaeckel:2010ni,Ringwald:2012hr,Arias:2012az,Graham:2015ouw,Marsh:2015xka,Irastorza:2018dyq,DiLuzio:2020wdo,OHare:2024nmr}, as well as~\cite{Cirelli:2024ssz} for generic dark matter).\footnote{From the theory side, a string axion with keV mass is a possibility because the accessible decay constant is around the string scale. Although the abundance from the standard misalignment mechanism~\cite{Preskill:1982cy,Abbott:1982af,Dine:1982ah} is typically too high, in the stochastic scenario it can be consistent the measured one with an electroweak Hubble scale inflation~\cite{Guth:2018hsa,Graham:2018jyp,Ho:2019ayl}.  Thermal production or misalignment production of anomaly-free ALP also matches the parameter region~\cite{Nakayama:2014cza,Takahashi:2020bpq,Sakurai:2022roq}. CP-even ALPs~\cite{Sakurai:2021ipp,Haghighat:2022qyh} also provide an option, potentially accessible via searches for the Higgs boson invisible decay at collider experiments. More recently, a feebly coupled Peccei-Quinn (PQ) model for the QCD axion, motivated by naturalness and the quality problem~\cite{Yin:2024txg}, predicts a light PQ Higgs boson as a dark matter candidate, which can be probed by the X-ray observation. Sterile neutrino dark matter can be around $\KEV$ mass range in the so-called Shi-Fuller mechanism~\cite{Shi:1998km} (see also Dodelson-Widrow mechanism ~\cite{Dodelson:1993je}).
The parameter region of the Dirac-type sterile neutrino dark matter with a large asymmetry may aslo be linked to the homochirality of the amino acid~\cite{Yin:2024trc}.}

Spectrographs with a resolving power of 
\[
R\equiv \l/\Delta\l=\O(1000)
\]
have recently become available or are on the verge of deployment. For instance, instruments such as Resolve on the X-Ray Imaging and Spectroscopy Mission (XRISM)~\cite{XRISMScienceTeam:2020rvx,Dessert:2023vyl}, the NIRSpec on the James Webb Space Telescope~\cite{jakobsen2022near,Bessho:2022yyu,Roy:2023omw}, 
MUSE at Very Large Telescope~\cite{bacon2017muse,Regis:2020fhw,Todarello:2023hdk}, and the Prime Focus Spectrograph (PFS) on the Subaru Telescope~\cite{PFSTeam:2012fqu} are designed to capture fine spectral details. On the other hand, the typical velocity dispersion of Milky Way dark matter is around 100~km/s, which determines the expected line broadening. This implies that the line from the dark matter two-body decay is broadened by the Doppler shift, with a narrow line width of roughly a factor of $\O(1/1000)$ (which coincides with $1/R$) of the line’s peak energy. Since a resolution larger than $R= \O(1000)$ would reduce the differential photon flux per spectral bin, while a resolution smaller than $R=\O(1000)$ would mean that each bin captures only a smaller fraction of the dark matter signal, the optimal choice is $R\sim 1000$ for searching for the dark matter line from the Milky Way. For a dark matter search, other than the Milky way center, one also observes dark matter--rich galaxies such as dwarf spheroidal galaxies (dSphs) and galaxy clusters, which have velocity dispersions smaller than 10~km/s and larger than 500~km/s, respectively.\footnote{If the field-of-view is very small and one observes the center of the galaxy, the differential $D$-factor of dSphs can be significantly enhanced depending on the inner profile~\cite{Yin:2023uwf,Bessho:2022yyu}. An optimized detector for the dSphs are the high dispersion spectrograph with $R\gtrsim 10^4$ (see \cite{Bessho:2022yyu,Yin:2024lla,Bessho:2024tpl}). In this case, the Milky way component gets suppressed in each bin. } Since these targets are not optimized for $R\sim 1000$, even though they are dark matter--rich, a line-of-sight observation of an extragalactic target inherently includes a significant contribution from Milky Way dark matter. Given the sizable radial velocities of many targets and the high resolution of the spectrographs, we point out that {\it the dark matter signals appear as two (or even more) narrow lines in various observations}. This superposition of signals is predicted to yield a two-peak structure in the observed spectrum—a feature that could serve as a distinctive signature of dark matter decay if properly disentangled.

In fact, one does not necessarily need to specifically target dark matter searches; observations conducted for other purposes may also be used for dark matter searches.

The XRISM Collaboration~\cite{XRISM:2025axf}
presents high-resolution X-ray spectroscopic observations of the Centaurus cluster core using the Resolve instrument with $R\approx 1500$. Their analysis reveals that the hot intracluster medium exhibits coherent bulk motion—with line-of-sight velocities ranging from about 130 to 310\,km\,s$^{-1}$ within roughly 30\,kpc of the center—and low turbulent velocity dispersions ($<120$\,km\,s$^{-1}$), findings that support a scenario in which gas sloshing, rather than vigorous AGN-induced turbulence, redistributes thermal energy to counteract radiative cooling.

In this paper, we use the same XRISM data to search for dark matter by taking into account the two narrow lines, assuming that the background spectrum is a continuous function after masking several lines consistent with the energy of known atomic lines. We also perform the conventional single-narrow-line analysis and show that the two-line analysis can remove many highly significant “signals”, thereby demonstrating the power of our approach.

On the other hand, the single-line constraint is also useful in some scenarios in which the photon is more likely to originate from a distant galaxy cluster rather than the Milky Way or nearby smaller galaxies. This is the case for hot dark matter from heavy particle decays (e.g., reheating~\cite{Jaeckel:2020oet,Jaeckel:2021gah,Jaeckel:2021ert}) or cold dark matter decaying into long-lived mediators which convert to photons.

From a $\chi^2$ fit, we derive the 2$\sigma$ limit for the dark matter lifetime. Due to the short observation time, for cold dark matter our limit is typically weaker than some existing limits; however, in the search for distant galaxy clusters our limit is among the strongest (at least in certain mass ranges), and thus it can constrain the exotic scenarios discussed above.

This result underlines the potential of future, more extensive observational campaigns to further constrain dark matter properties through spectroscopy.

This paper is organized as follows. In \Sec{2}, we review the estimation of the photon flux from the two-body decay of dark matter and point out the prediction of the two narrow lines. In Section~\Sec{3}, we demonstrate the dark matter search analysis using the recent XRISM data and set a limit on the dark matter lifetime. The final section (\Sec{conclusion}) is devoted to conclusions and discussion. 

We adopt natural units in this paper. For example, a velocity expressed without an explicit unit is understood to be normalized by the speed of light.

\section{Precise Photon Spectra from Dark Matter Decay}\lac{2}
Let us consider the decay of dark matter into two particles, one of which is a photon. In the case of a sterile neutrino, only one photon is produced, whereas for an ALP, two photons are produced. To study this system in a general manner, we introduce the dark matter mass $m_\phi$, the decay rate $\Gamma_\phi$ ($\Gamma^{-1}_\f$ is the lifetime), and the number of photons $q$ (with $q=1$ or 2) produced in each reaction.

The resulting differential flux of photons from the decay consists of two components,
\beq
\frac{d\F_\gamma}{dE_\gamma} = \frac{d\F_\gamma^{\text{extra}}}{dE_\gamma} + \sum_i \frac{d\F_{\gamma,i}}{dE_\gamma}\,.
\eeq
Here the first term represents the extragalactic component, which is isotropic and has a continuous spectrum. We do not consider it since the event rate for this component within a spectral bin is suppressed by the large $R$ (and in the analyses we consider in \Sec{3} this component is subtracted).

Our task is to estimate the second term, which arises from localized contributions such as those from our Milky Way or target galaxies for the observations. Here $i$ denotes the galaxy (cluster). This term can be expressed as
\beq
\frac{d\Phi_{\gamma,i}}{dE_\gamma} = \int_{\D\Omega} ds\, d\Omega\, \frac{e^{-\tau[s,\Omega]\, s}}{4\pi s^2} \left(\frac{\Gamma_\phi\, \rho^{i}_\phi(s,\Omega)}{m_\phi}\right) s^2\, \frac{dN_{\phi,i}}{dE}[s,\Omega]\,,
\eeq
where $s$ is the line-of-sight distance, $\Omega$ is the solid angle, 
and the integration range of the solid angle, $\Delta\Omega$, corresponds to the field-of-view of the detector.  
The other quantities and their calculation are detailed below. 

\begin{itemize}
\item[$\bullet$] $\tau$ is the (averaged) optical depth. We take
\beq
\tau \simeq 0
\eeq
for simplicity in this paper.

\item[$\bullet$] $\rho^{i}_\phi$ represents the dark matter density distribution, which depends on the dark matter halo model and the properties of galaxy $i$. For instance, considering the Navarro-Frenk-White (NFW) profile~\cite{Navarro:1996gj}, we have
\beq\laq{NFW}
\rho^i_\phi(r) = \frac{\rho_{0,i}}{\displaystyle \frac{r}{r_{s,i}}\left(\frac{r}{r_{s,i}}+1\right)^2}\,,
\eeq
with $r$ being the distance from the center of the galaxy. The parameters $r_{s,i}$ and $\rho_{0,i}$ can be obtained from observational quantities such as rotation curves and the temperature distribution in the intracluster medium.

\item[$\bullet$] $\frac{dN_{\phi,i}}{dE}[s,\Omega]$ represents the photon spectrum from a single dark matter decay in the vicinity of galaxy $i$. In particular, the photon spectrum is Doppler-shifted as
\beq
\frac{dN_{\phi,i}}{dE} = q \int d^3\vec{v}\; f_i(v,s,\Omega)\; \delta\Big(E-(1-\vec{v}\cdot\vec{\Omega})\frac{m_\phi}{2}\Big)\,.
\eeq
Here $f_i$ is the dark matter velocity distribution at $(s,\Omega)$, and the factor $q$ ensures that the integral over $E$ yields the average number of final photons. Throughout this paper, we assume that the dark matter velocity follows a symmetric Gaussian distribution,
\beq\laq{fdis}
f_i = \prod_{d=1}^3 \frac{1}{\sqrt{2\pi}\, \s_i} \exp\Big[-\frac{([\vec{v}-\vec{v}_i]_d)^2}{2\s_i^2}\Big]\,,
\eeq
where $\s_i$ is the one-dimensional velocity dispersion and $\vec{v}_i$ is the average velocity of galaxy $i$ in the detector frame. We assume the non-relativistic limit and perform a Galilean transformation. Then we obtain
\beq\laq{spectrum}
\frac{dN_{\phi,i}}{dE} \simeq \frac{2q}{m_\phi} \frac{1}{\sqrt{2\pi}\, \s_i}\exp\Big[-\frac{\Big(\frac{2E}{m_\phi}-1+\vec{v}_i\cdot\vec{\Omega}\Big)^2}{2\s_i^2}\Big] \approx \frac{2q}{m_\phi} \frac{1}{\sqrt{2\pi}\, \s_i}\exp\Big[-\frac{\Big(\frac{2E}{m_\phi}-1+v_{r,i}\Big)^2}{2\s_i^2}\Big]\,.
\eeq
In the last approximation we assume that the field-of-view is sufficiently small so that $\vec{\Omega}$ can be approximated the averaged one in the field-of-view. 
Thus $\vec{v}_i\cdot\vec{\Omega}$ is replaced by the radial velocity $v_{r,i}$.

\item[$\bullet$] The velocity dispersion $\s_i$, which generally depends on $\vec{r}$, can also be obtained self-consistently. Assuming $\s_i[\vec{r}]=\s_i[r]$ for a spherical distribution around the galaxy center, the one-dimensional velocity dispersion is given by the Jeans equation~\cite{Robertson:2009bh} as
\beq\laq{dispersion}
\s_i^2[r] = \frac{1}{\rho_{\phi,i}[r]} \int_{r}^{\infty} \rho_{\phi,i}[r']\, \frac{d\F_{\rm grav}}{dr'}\,dr'\,,
\eeq
with $\F_{\rm grav}$ being the gravitational potential of the galaxy. Assuming that dark matter dominates the gravitational potential, this can be estimated using elementary Newtonian gravity. One finds that the velocity dispersion attains a maximum at $r\sim r_{s,i}$ and is suppressed for $r\gg r_{s,i}$ or $r\ll r_{s,i}$. For the Milky Way one typically has $\s_i(r_{s,i})\sim 100\,\rm km/s$, for a typical galaxy cluster $\s_i(r_{s,i})\gtrsim 500\,\rm km/s$ (e.g., the Centaurus cluster has $\s_i(r_{s,i})\sim 600\,\rm km/s$), and for a dSph we expect $\s_i(r_{s,i})\lesssim 10\,\rm km/s$.
\end{itemize}

In the limit where the velocity dispersion is much smaller than the energy resolution of the detector, we can approximate $\frac{dN_{\phi,i}}{dE}$ as a delta function. In that case, it becomes independent of $\s_i$ (and hence of $s$ or $\Omega$), allowing us to define the $D$-factor as
$
D_i \equiv \int_{\Delta\Omega} d\Omega\, ds\, \rho_\phi^i\,,
$
since the dependence on $s$ and $\Omega$ factorizes (see Eq.~\eq{spectrum}).  Note that the commonly used $D$-factor formalism cannot be applied straightforwardly when $\s_i$ is not negligible, because then $\frac{dN_{\phi,i}}{dE}$ depends on $s$ and $\Omega$ through $\s_i (s,\Omega)$.

Now we come to the key point of our study. Consider searching for the narrow-line photons from dark matter decays. For this purpose, we assume a detector with a sufficiently small field-of-view and high spectral resolution ($R \gg 1$); that is, we examine the photon deposit in each spectral bin over the range 
\[
\sim \Big[E_a\Big(1-\frac{1}{2R}\Big),\,E_a\Big(1+\frac{1}{2R}\Big)\Big],
\]
with $E_a$ being the energy of the $a$th bin. The differential photon flux is then measured as (photon counts in each bin) divided by ($E_a/R$).

As shown in, e.g., Refs.~\cite{Combet:2012tt,Evans:2016xwx}, the $D$-factors, the use of which implies the limit $\s_i\to 0$, of various dSphs or galaxy clusters can be higher than that of the Milky Way background—this is the reason why the Milky Way contribution is often neglected when searching for dark matter in other objects. In many cases, the extragalactic $D$-factor is higher by an order of magnitude.

On the other hand, differences in velocity dispersion, which broaden the photon spectrum via the Doppler shift, affect the signal in each spectral bin. According to the discussion below Eq.~\eq{spectrum}, the typical relative width of the narrow line is
\beq
\frac{\Delta_i E_{\rm signal}}{E_{\rm signal}} \sim \s_i\,,
\eeq
which is approximately $\sim 3\times 10^{-4}$ for the Milky Way, $\lesssim 3\times 10^{-5}$ for dSphs, and $\gtrsim 2\times 10^{-3}$ for galaxy clusters. This implies that for a spectral resolution of $R\sim 1000$—as implemented in instruments such as XRISM/Resolve, VLT/MUSE, JWST/NIRSpec, and the  Subaru/PFS—the resolution coincides with the Milky Way’s ratio $\frac{\Delta_{\rm Milky Way} E_{\rm signal}}{E_{\rm signal}}$, meaning that $R$ is optimized for photons from the Milky Way component.
 In contrast, the signal from a galaxy cluster appears fainter in each spectral bin by a factor of 
$
\frac{R^{-1}}{\Delta_{\rm cluster}E_{\rm signal}/E_{\rm signal}}.
$
The dark matter from the dSphs and Milky way component both looks like a narrow line. However, when the field-of-view is much larger than the dSphs, or when we look at some dSphs that do not have too many dark matter, the signal from dark matter decay from dSphs could also have a signal not much larger than the one from the Milky Way background. 

Thus, we claim that when using detectors with $R\sim1000$, it is necessary to account for the dark matter contribution from the Milky Way background. Interestingly, $R\sim 1000$ can already resolve various radial velocities—e.g., in most galaxy clusters and a large portion of dSphs such as dark matter-rich Draco or Segue~1. Therefore, when performing dark matter searches with these detectors, the dark matter signal could appear as double narrow lines (see, e.g., Figs.~\ref{fig:0} and ~\ref{fig:1}).\footnote{In some galaxy clusters the central galaxy may have a relative motion with respect to the cluster; in that case, three narrow lines may appear. In the case Centaurus galaxy cluster, we cannot find past studies suggesting the center galaxy, NGC4696, to give a significant enough contribution.  }

This effect could serve as a smoking gun signal for dark matter in these advanced detectors. Depending on the analysis, the multiple narrow lines must be taken into account; otherwise, one might erroneously reject a dark matter signal.

\section{Searching for Double Narrow Lines from XRISM Data}
\paragraph{Flux Data}
Now we perform a real data analysis in the search for dark matter. To this end, we employ the data from Ref.~\cite{XRISM:2025axf} obtained from observations of the Centaurus Cluster. Relevant parameters for the calculations are listed in Table~\ref{tab:1}.

\begin{table}[htbp]
\centering
\caption{Relevant parameters for estimation. These are obtained from Refs.~\cite{XRISM:2025axf} and Refs.~\cite{Cautun:2019eaf,Combet:2012tt} for the dark matter profiles. Coordinate is taken from  SIMBAD database~\cite{Wenger:2000sw}. The radial velocity is taken from Refs.~\cite{XRISM:2025axf}. }
\begin{tabular}{ll}
\toprule
\textbf{Parameter} & \textbf{Value} \\[2mm]
\midrule
Instrument & XRISM/Resolve \\[2mm]
Observation Period & December 2023 -- January 2024 \\[2mm]
Total Exposure Time & 284.7\,ks \\[2mm]
Full field-of-view (FoV) & $2.9'\times2.9'$ (excluding the southwest block)~\cite{XRISM:2025axf} \\[2mm]
Aim Point (Galactic Coordinate, J2000) & $\ell_a\approx 302.396^\circ,\; b_a\approx 21.569^\circ$ \\[2mm]
Photon Energy Range & 1.8--8\,keV \\[2mm]
\bottomrule
Target~1 & Centaurus Cluster (A3526) \\[2mm]
Center Coordinate & $\ell\approx 302.403^\circ,\; b\approx +21.564^\circ$ \\[2mm]
Distance to Center & 48.1\,Mpc \\[2mm]
$\rho_{s}$ & $0.025\,\GEV\,\rm cm^{-3}$ \\[2mm]
$r_{s}$ & 0.33\,Mpc \\[2mm]
Radial Velocity $v_r/c$ & 0.0104 \\[2mm]
\bottomrule
Target~2 & Milky Way Background \\[2mm]
Center Coordinate & $\ell = 0^\circ,\; b=0^\circ$ \\[2mm]
Distance to Center & 0.0082\,Mpc \\[2mm]
$\rho_{s}$ & $0.46\,\GEV\,\rm cm^{-3}$ \\[2mm]
$r_{s}$ & 0.0144\,Mpc \\[2mm]
Radial Velocity (along the line-of-sight) $v_r/c$ & $220\,\rm km/s\,\cos(90^\circ-\ell_{a})\cos(b_a)\approx 0.00058$ \\
\bottomrule
\end{tabular}
\label{tab:1}
\end{table}

The measured photon flux is obtained from the count data, $d[E_a]$, in units of (counts\,s$^{-1}$\,keV$^{-1}$) for the $a$th bin (as measured by Resolve in \cite{XRISM:2025axf}), and the effective area, $A[E]$, in units of cm$^2$ (from Fig.~14 of 
{\url{https://xrism.isas.jaxa.jp/research/analysis/manuals/xrqr_v1.pdf}}
). The measured differential photon flux is then given by
\beq
F[E] = \frac{d[E]}{A[E]}\,,
\eeq
in units of cm$^{-2}$\,s$^{-1}$\,keV$^{-1}$. We can derive the data points $F[E_a]$, with a 1$\sigma$ error $\Delta F[E_a] = \Delta d[E_a]/A[E_a]$, as well as the bin half-width $\Delta E_a$ for the $a$th bin. 

{Then, we mask some known emission lines, as shown in Table \,\ref{tab:2}. 
The line profiles and energies are obtained from the atomDB database~\cite{Smith:2001he,Foster:2012hy}.\footnote{\url{http://www.atomdb.org}}.
For the lines expected from NGC 4696 ($v_r=0.01003$), the central galaxy of A3526, we take the Doppler shift into account.
For line number 17, the line is narrower than the others, and a mask width of $0.0015$ is sufficient.\footnote{Although this may imply that this is a Ca XIX line from the Milky Way, the line energy is also consistent with that of Ar XVIII from NGC 4696.}
}

Before going into details, let us provide some sources of systematics.
{Although we only used the effective area for the estimation, the response function of Resolve includes not only the effective area but also the precision in determining the energy centroid of emission lines (gain) and the line-spread function. Since there may be uncertainties in these components, further investigation is necessary.} {We also note that the dark matter profile itself is subject to systematic uncertainties and may vary. For instance, if we consider an asymmetric DM profile--such as a triaxial halo model consistent with observations--even the well-known Milky Way component could be affected. On the other hand, the local enhancement due to the dark matter of NGC 4696, as opposed to that of A3526, could also improve dark matter detection. Investigating these systematic uncertainties in more detail will be the focus of our future work.}

\begin{figure}[!t]
\begin{center}  
   \includegraphics[width=170mm]{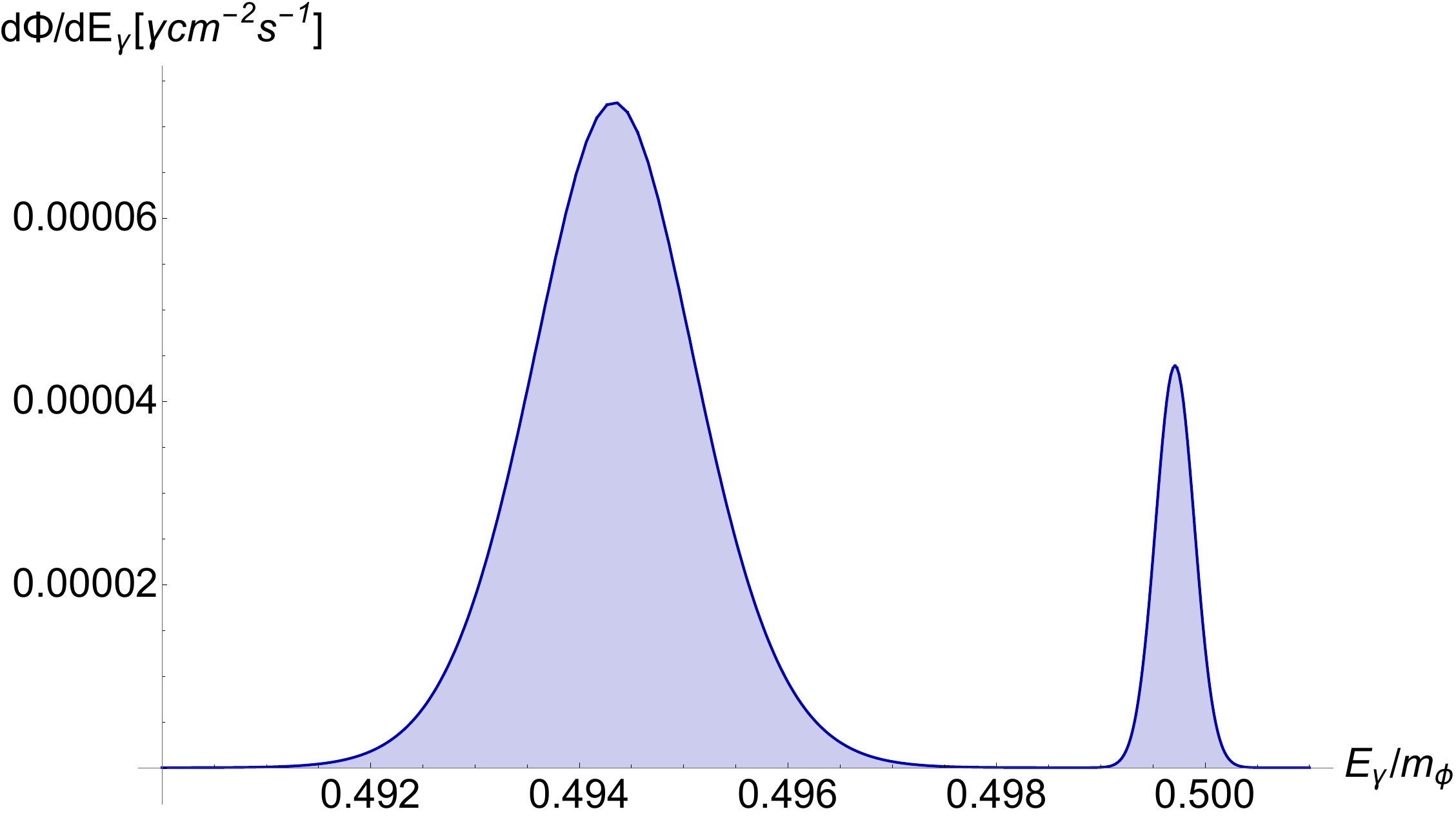}
\end{center}
\caption{
The dark matter signal flux, $d\F/dE$, derived using the parameters in Table~\ref{tab:1}. Here we use $q\Gamma_\phi=2\times 10^{-19}\,\mathrm{yr}^{-1}$ and $m_\phi=10\,\mathrm{keV}$. In general, $d\F/dE\propto q\Gamma_\phi/m_\phi^2$.
} \label{fig:0}
\end{figure}

\begin{table}[htbp]
  \centering
  \caption{Applied mask. }
  \label{tab:2}
  \begin{tabular}{cccc}
    \hline
    No. & Energy (keV) & Element Composition & Total width \\
    \hline
    1  & 1.865  & Si~XIII  & 0.015 \\
    2  & 2.006  & Si~XIV   & 0.015 \\
    3  & 2.377  & Si~XIV   & 0.015 \\
    4  & 2.430  & S~XV     & 0.015 \\
    5  & 2.461  & S~XV     & 0.015 \\
    6  & 2.506  & Si~XIV   & 0.015 \\
    7  & 2.622  & S~XVI   & 0.015 \\
    8  & 2.884  & S~XV     & 0.015 \\
    9  & 3.107  & S~XVI   & 0.015 \\
    10 & 3.124  & Ar~XVII  & 0.015 \\
    11 & 3.140  & Ar~XVII  & 0.015 \\
    12 & 3.320  & Ar~XVIII & 0.015 \\
    13 & 3.276  & S~XVI   & 0.015 \\
    14 & 3.861  & Ca~XIX  & 0.015 \\
    15 & 3.888  & Ca~XIX  & 0.015 \\
    16 & 3.902  & Ca ~XIX& 0.015 \\
    17 & 3.902  & Ca~XIX (MW) & 0.0015 \\
    18 & 4.107  & Ca~XX   & 0.015 \\
    19 & 6.655  & Fe & 0.2  \\
    20 & 6.700  & Fe~XXV  & 0.03  \\
    21 & 7.882  & Fe~XXV  & 0.03  \\
    \hline
  \end{tabular}
\end{table}

\paragraph{Signals from Standard Cold Dark Matter}
To compare with the potential dark matter signals, we estimate the signal flux according to the discussion in the previous section without employing the $D$-factor formalism. Here, we integrate the dark matter profiles over the field-of-view of Resolve (excluding the southwest block not used for data taking) for $i=\rm Centaurus~Cluster$ and for the Milky Way background.\footnote{For the Centaurus cluster, we did not shift by $1'$ from the central axis; the effect of such a shift was checked to be an $\mathcal{O}(1)\%$ correction to the flux.} We note that the radial velocity of the Milky Way component is not zero because the observation is offset from the Galactic center. Taking into account the Doppler effects from radial velocities and velocity dispersions, we obtain the differential flux $d\F_\gamma/dE_\g$ from dark matter decays for fixed values of the mass and $q\Gamma_\phi$. This is displayed in Fig.\,\ref{fig:0} with respect to $E_\g/m_\f$. We take $q\G_\f=2\times 10^{-19}\rm yr^{-1}$ and $m_\f=10\rm keV$. From the discussion in \Sec{2}, one can easily find $d\F/dE_\g\propto q \G_\f /m_\f$.

Our data analysis is performed as follows (see e.g. \cite{Kaastra:2006pq}). We fit the relevant data with a cubic spline function,
$
f(E)=\sum_{n=0}^3 c_n E^n\,,
$
together with the dark matter signal $d\F/dE$, so that the total flux is
$
J_{\rm tot}= \frac{d\F_\gamma}{dE}+f(E)\,.
$
Fixing the dark matter mass, we have four parameters for the continuous function and one parameter, $q\Gamma_\phi\,(>0)$, at this stage. Then, over a certain energy range, we define
$
\chi^2 = \sum_{a}\Delta F[E_a]^{-2}\left(\frac{1}{2\Delta E_a}\int_{E_a-\Delta E_a}^{E_a+\Delta E_a}dE\, J_{\rm tot} - F[E_a]\right)^2\,.
$
To search for dark matter, we estimate the best-fit difference
$
\Delta\chi^2 \equiv \min_{(c_n)}\chi^2\Big|_{\Gamma_\phi\to0} - \min_{(c_n,\,q\Gamma_\phi)}\chi^2\,,
$
where the second term is minimized with respect to both the coefficients $c_n$ and $q\Gamma_\phi^0$. The significance for the dark matter hypothesis relative to the background hypothesis (with $q\Gamma_\phi^0$) is then given by $\sqrt{\Delta\chi^2}$.
In the fit we do not include the masked data points. In addition, we do not consider the case that either narrow line peak is in the masked region.

For a given dark matter mass, we select an energy bin spanning 5 times the energy difference between the two narrow line peaks around the average energy.\footnote{A larger range would provide a more stringent dark matter limit. } 
{An example fit is shown in Fig.\ref{fig:1} in the red solid line. The black points are the flux data points. The blue dashed line represents the 2$\s$ limit for the dark matter life time discussed soon. 
This best fit (red solid line) has a local significance $\sim 2.3\s$ over the absence of the dark matter. 

Given that there should be some weak emission lines that we have not yet masked, it should be necessary to subtract the sky background or to model and subtract the baryonic (or instrumental) background. The former approach, using a sky region behind the Galactic center to suppress the Milky Way component, is discussed in~\cite{Foster:2021ngm} in the context of blank sky search. If the sky region around the target galaxy is used, the Milky Way component would be subtracted; in such observations, one should employ the single-peak analysis. 

We found an approximate $5\sigma$ local significance near $m_\f = 15.6,\mathrm{keV}$ for the narrow line from the galaxy cluster; however, the region is close to unmasked lines from Fe or Ni. On the other hand, when neglecting the Milky Way contribution, the significance increases to approximately $6$--$7\sigma$. Thus, in the two-narrow-line analysis, we successfully reduced the impact of these highly significant emission lines. In a more sophisticated analysis—first subtracting the background via sky subtraction or modeling baryonic lines and then performing the two-narrow-line fit—our approach should be even more powerful for detecting dark matter, provided that the dark matter distribution is well understood.

\paragraph{Dark Matter Limit}\lac{3}
We can also set limits on the dark matter lifetime.
To this end, we solve
\[
\min_{(c_n)}\chi^2[q\Gamma_\phi]-\min_{(c_n,q\Gamma_\phi)}\chi^2=4
\]
to obtain the 2$\sigma$ bound for $q\Gamma_\phi$ (with $q\Gamma_\phi>q\Gamma_\phi^0\ge0$). The result is shown in Fig.~\ref{fig:2} (left panel for the double narrow-line constraint and right panel for the single-peak constraint). For comparison, we also indicate the constraint from XMM-Newton blank-sky 547\,Ms data as derived in~\cite{Foster:2021ngm} in the left panel. 

\begin{figure}[!t]
\begin{center}  
 \includegraphics[width=140mm]{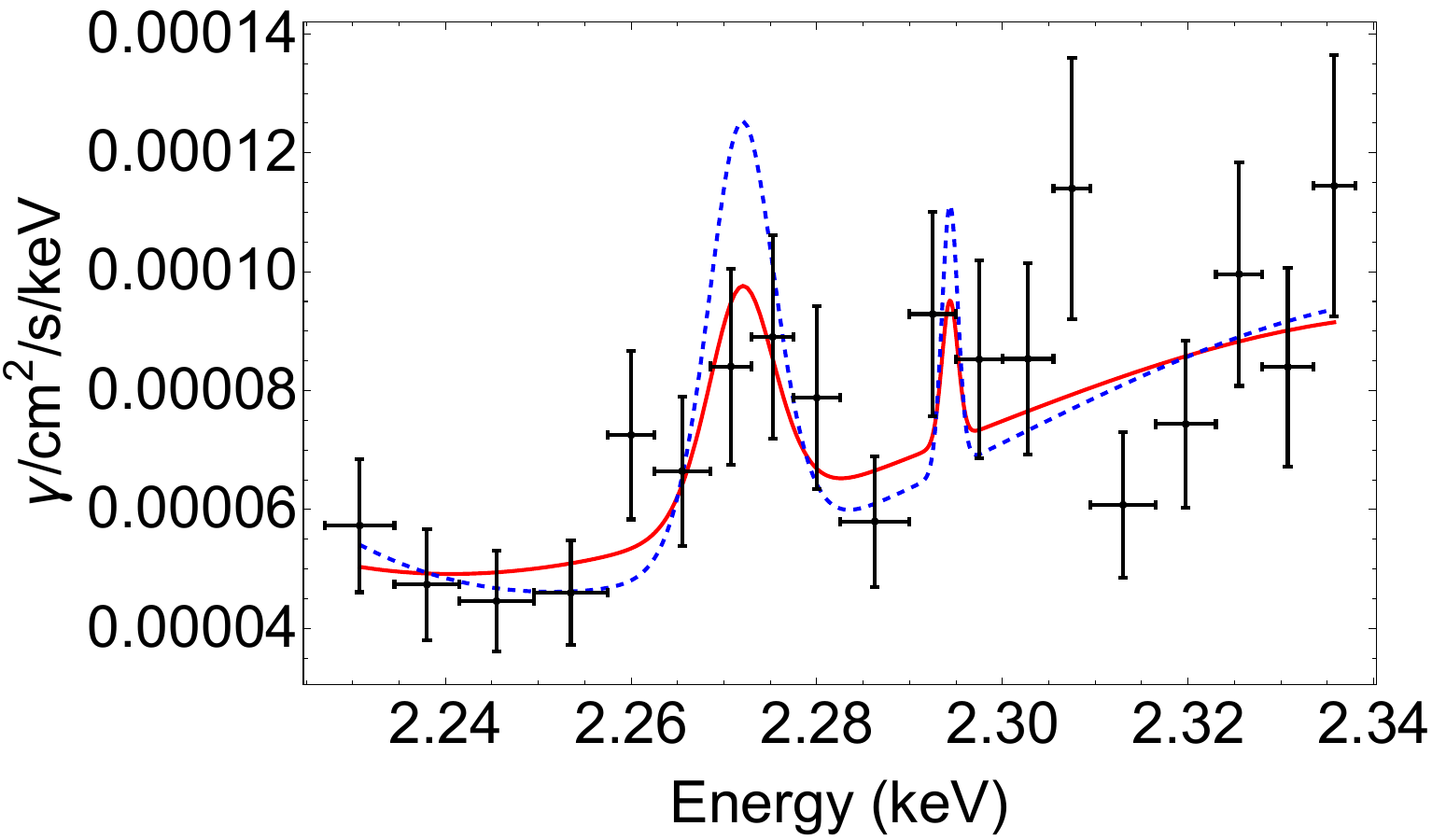}
      \end{center}
\caption{
A sample point for the best fit (red solid line) and 2$\s$ limit of the flux (blue dashed line). 
Here the best fit corresponds to $m_\phi\approx 4.60\KEV, 1/(q\Gamma_\f)\approx 8.8\times 10^{19}$yr which is preferred by $2.3\sigma$ level compared to $\G_\f\to 0.$ The 2$\s$ limit corresponds to $1/(q\G_\f)\approx 4.7\times 10^{19}$yr.
} \label{fig:1}
\end{figure}

On the other hand, some beyond Standard Model (BSM) particles may provide photons mostly from the galaxy cluster rather than the Milky Way. Two concrete scenarios are as follows:
\begin{itemize}
 \item \textbf{Hot DM from Inflaton/Moduli early Decays, Heavy Particle Decays, or Phase Transitions.~\cite{Jaeckel:2020oet,Jaeckel:2021ert,Jaeckel:2021gah,Azatov:2020ufh,Azatov:2021ifm,Baldes:2022oev,Azatov:2024crd,Ai:2024ikj}\footnote{Thermal freeze-in production may also contribute. Although the keV scale is relatively heavy and the typical post--galaxy-formation velocity is too slow for dark matter to escape the Milky Way’s gravitational potential, in some models with higher-dimensional operators a high-velocity component can be produced thermally~\cite{Sakurai:2021ipp}. }
} 
In this case, the BSM particle cannot be the dominant dark matter component if its free-streaming length is too long, but it can be a subdominant component. If the free-streaming length is long—implying that the typical velocity around and after galaxy formation is too high for dark matter to be captured by small galaxies with low escape velocities—it may still be captured by a galaxy cluster with a sizable escape velocity~\cite{Jaeckel:2021gah} (see also a recent detailed study~\cite{Dror:2024ibf}). In the galaxy cluster, the usual cold dark matter coexists with a fraction of the trapped BSM particle. We assume that the BSM particle follows the same distribution as dark matter (i.e., it is in Jeans equilibrium). This scenario predicts a single narrow-line photon. We denote the fraction by $r\, (<1)$.

 \item \textbf{Decay of Cold Dark Matter into a Pair of Mediators.} One may also consider the decay of cold dark matter into a pair of mediators that subsequently convert into photons. A simple example is the decay of cold dark matter into a pair of ALPs, which then convert into photons due to the extragalactic magnetic field.\footnote{Alternatively, one may consider mediators decaying directly into photons; in that case, the analysis must take into account the spectral shape.} This process produces a photon spectrum identical to that of the two-body decay. The conversion may occur efficiently over distances of order $\mathcal{O}$(Mpc), depending on the ALP-photon coupling and the magnetic field strength. Thus, while the Milky Way or nearby galaxies do not contribute significantly, a distant galaxy cluster contributes to the production of a single narrow-line photon. We denote the conversion rate by $r\,(\leq 1)$.
\end{itemize}

In these cases, the limits from observations of the Milky Way or nearby galaxies/dSphs are alleviated, while the limits from galaxy clusters remain important. In a model-independent manner, we show the limit on $r\,q\Gamma_\phi$ in the right panel of Fig.~\ref{fig:2}, assuming a single narrow-line from dark matter decay in the Centaurus Cluster while neglecting the Milky Way component. For comparison, we also show the limits from Suzaku 537\,ks data of the Perseus Cluster in light gray region~\cite{Tamura:2014mta}\footnote{{See \cite{2024PASJ76512F} for more recent study and also, e.g., Refs.\,\cite{Boyarsky:2014jta,Bulbul:2014sua,Boyarsky:2014ska,Urban:2014yda} for the 3.5keV line excess. }} and NuSTAR 266\,ks data of the Bullet Cluster~\cite{Riemer-Sorensen:2015kqa}.

\begin{figure}[!t]
\begin{center}  
   \includegraphics[width=75mm]{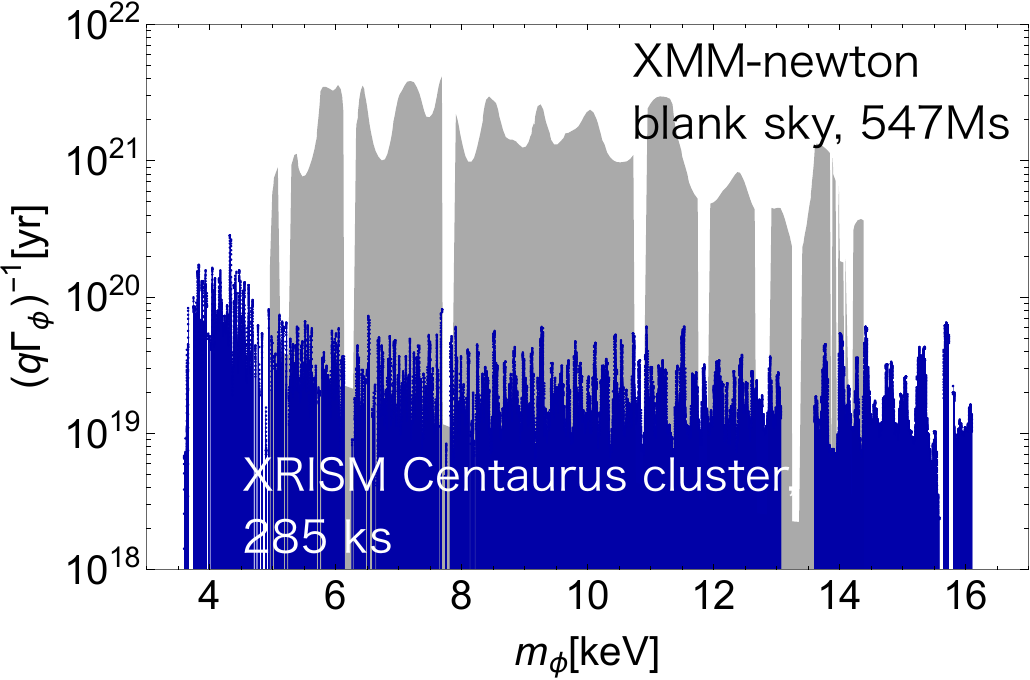}
~~~      \includegraphics[width=75mm]{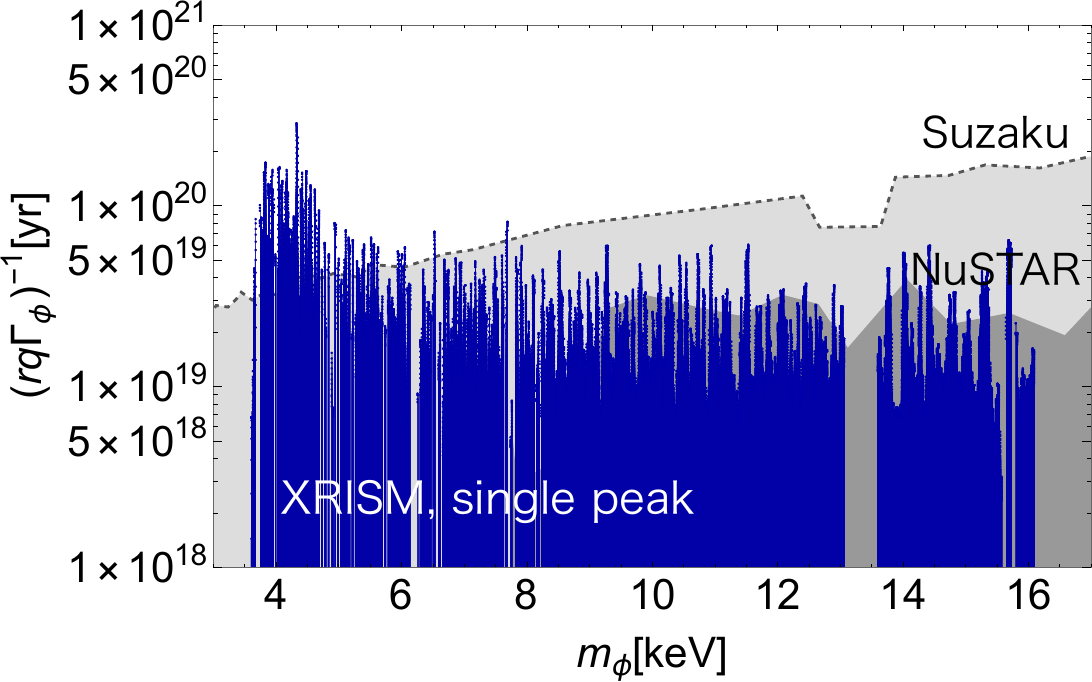}
\end{center}
\caption{
Left panel: The 2$\sigma$ limits on the double-peak structure from dark matter decay derived in this paper. Also shown, for comparison, is the limit from XMM-Newton blank-sky 547\,Ms data as derived in~\cite{Foster:2021ngm}. 
Right panel: The limit for the single peak from the Centaurus Cluster, which can be applied to hot dark matter that is only captured in galaxy clusters. For comparison, we also show the bounds from Suzaku 537\,ks data of the Perseus Cluster~\cite{Tamura:2014mta} and NuSTAR 266\,ks data of the Bullet Cluster~\cite{Riemer-Sorensen:2015kqa}.
} \label{fig:2}
\end{figure}

\section{Conclusions}\lac{conclusion}
In this paper, we have investigated the two-body decay of dark matter particles into final states that include narrow photon lines, a process naturally arising in various dark matter models such as those involving axion-like particles and sterile neutrinos. We have demonstrated that the Doppler shift effects from both the Milky Way background and various extragalactic targets lead to a distinctive double narrow-line signature in the spectrograph with $R\sim 1000$.

The two-narrow-line fitting approach has proven effective in mitigating the contamination from baryonic emission lines, thereby enhancing the sensitivity to potential dark matter signals. This study also underscores the necessity of accounting for the Milky Way dark matter contribution when analyzing extragalactic targets, as its impact becomes significant at the optimized spectral resolution of $R\sim 1000$.

Our analysis of recent XRISM data for the Centaurus Cluster has yielded constraints on the dark matter decay rate that are competitive or stronger in certain mass ranges when compared to existing limits for galaxy clusters. However it is typically weaker than the strongest bound from the blank sky search for the Milky way dark matter. That said our limit can be useful probing a variety of beyond the Standard Model scenarios that induce the narrow line photon from distant galaxy clusters due to the short integration time.  

Overall, our work demonstrates the promising potential of high-resolution X-ray spectroscopy as a powerful tool in the indirect search for dark matter.
Future observations with longer exposure times and improved background subtraction techniques will be crucial to further tighten these constraints.

\section*{Acknowledgement}
This work was supported by JSPS KAKENHI Grant Nos.  20H05851 (W.Y.),  22K14029 (W.Y.), and 22H01215 (W.Y.), JP22H01268(Y.F.), JP22H00158(Y.F.), JP22K03624(Y.F.), JP23H04899 (Y.F.). 
W.Y. is also supported by Incentive Research Fund for Young Researchers from Tokyo Metropolitan University. 

\appendix 

\section{Limits on ALP and Sterile Neutrino Parameters}

The ALP lifetime is given by 
\beq
\G_\f^{-1} \approx 4.2\times 10^{24} {\rm yr} \(\frac{10^{18}\GEV}{g_{\f\g\g}}\)^2\(\frac{\KEV}{m_\f}\)^3
\eeq
with $g_{\f\g\g}$ being the ALP--photon coupling. (Here $q=2$.)

The sterile neutrino lifetime is given by 
\beq
\G_\f^{-1}\approx 2.3\times 10^{24} {\rm yr} \frac{10^{-10}}{\sin^2(2\theta)} \(\frac{\KEV}{m_\f}\)^5.
\eeq
with $\sin^2(2\theta)$ being the active--sterile neutrino mixing angle. (Here $q=1$.)

Then, we can recast the lifetime limits from Fig.~\ref{fig:2} in terms of these parameters, as shown in Fig.~\ref{fig:para}.
\begin{figure}[!t]
\begin{center}  
   \includegraphics[width=80mm]{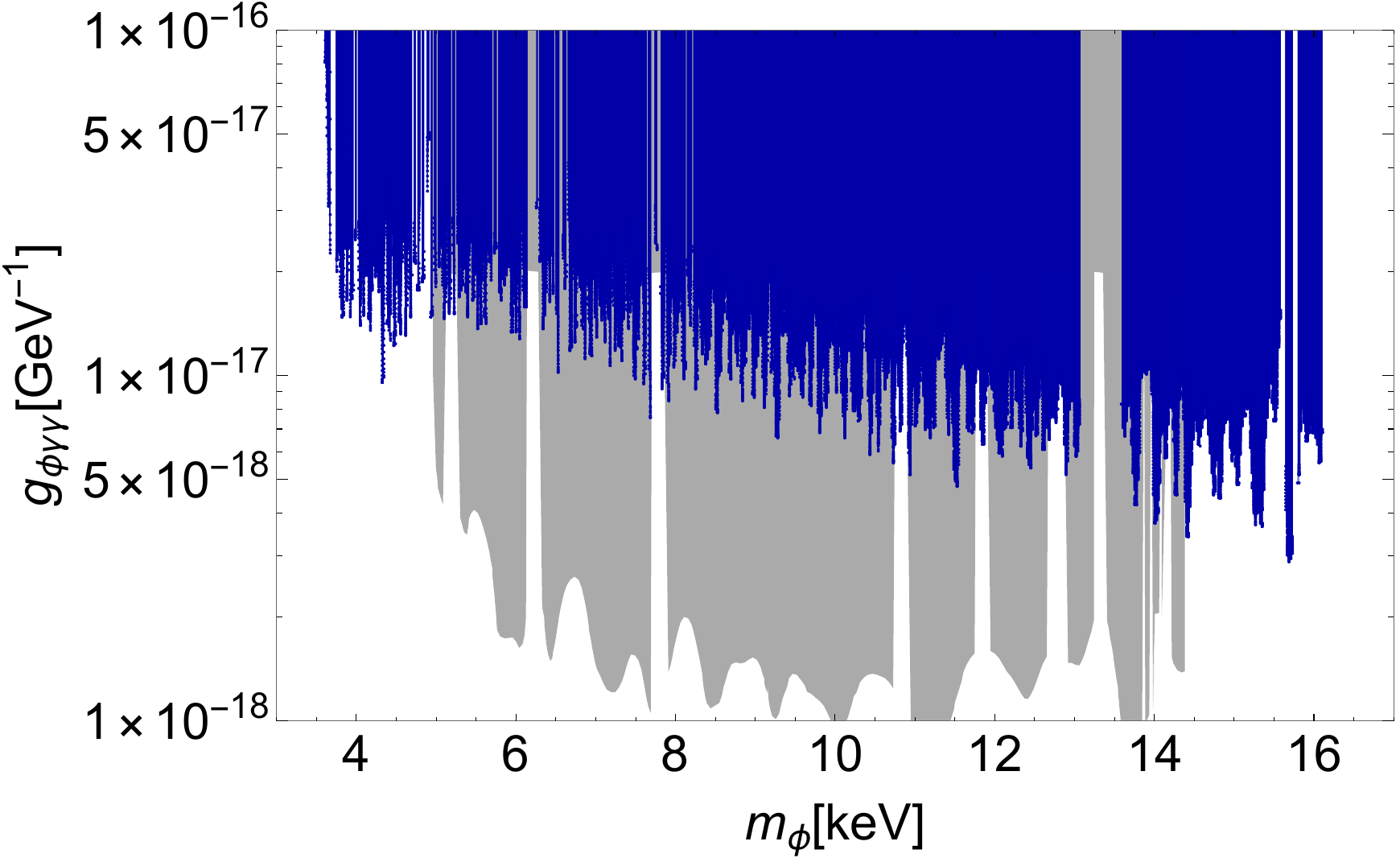}
   \includegraphics[width=75mm]{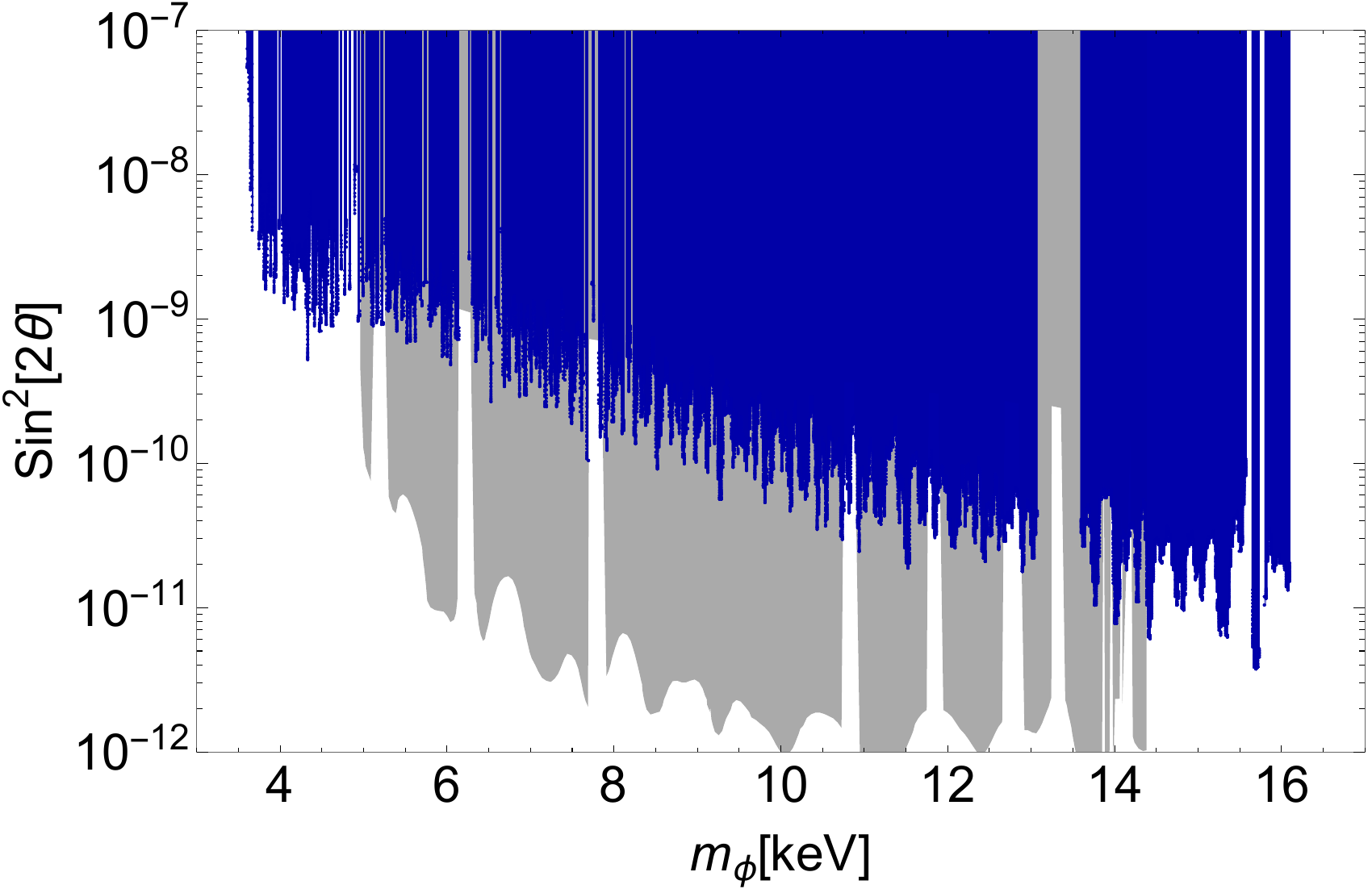}
\end{center}
\caption{
Same as the left panel of Fig.~\ref{fig:2}, but expressed in terms of the ALP--photon coupling (left panel) and the active--sterile neutrino mixing parameter (right panel).
} \label{fig:para}
\end{figure}

\bibliography{Xrism}

\end{document}